\documentclass[RNAAS]{aastex62}

\graphicspath{{./}{figures/}}

\begin{document}

\title{Eccentricity is Not Responsible for Odd Harmonics in HAT-P-7 and Kepler-13A}


\author{Claudia I. Bielecki}
\affiliation{Department of Physics, McGill University, 3600 rue University, Montr\'{e}al, QC Canada H3A 2T8}
\affiliation{McGill Space Institute, 3550 rue University, Montr\'{e}al, QC, H3A 2A7, CAN.}
\affiliation{Institut de recherche sur les exoplan\`{e}tes, Universit\'{e} de Montr\'{e}al, C.P. 6128, Succ. Centre-ville, Montr\'{e}al QC H3C 3J7, CAN}

\author{Nicolas B. Cowan}
\affiliation{Department of Physics, McGill University, 3600 rue University, Montr\'{e}al, QC Canada H3A 2T8}
\affiliation{Department of Earth \& Planetary Sciences, McGill University, 3450 rue University, Montr\'{e}al, QC Canada, H3A 0E8}
\affiliation{McGill Space Institute, 3550 rue University, Montr\'{e}al, QC, H3A 2A7, CAN.}
\affiliation{Institut de recherche sur les exoplan\`{e}tes, Universit\'{e} de Montr\'{e}al, C.P. 6128, Succ. Centre-ville, Montr\'{e}al QC H3C 3J7, CAN}

\email{claudia.bielecki@mail.mcgill.ca, nicolas.cowan@mcgill.ca}


\section{} 

\vspace{-1.1cm}
The exquisite photometry of \emph{Kepler} has revealed reflected light from exoplanets, tidal distortion of host stars and Doppler beaming of a star's light due to its motion \citep{2016RPPh...79c6901B,2012ApJ...751L..28D, 2010ApJ...713L.145W, 2012MNRAS.422.2600B}. 
\cite{2013ApJ...772...51E, 2015ApJ...804..150E} and \cite{2014ApJ...788...92S} reported additional odd harmonics in the light curves of two hot Jupiters: HAT-P-7b and Kepler-13Ab. They measured non-zero power at three times the orbital frequency that  persisted while the planet was eclipsed and hence must originate in the star \citep{2015ApJ...804..150E}. 
\cite{2018arXiv180307078P} showed that orbital eccentricity could result in time-dependent tidal deformation of the star that manifests itself at three times the orbital frequency and suggested this could be the origin of the measured odd modes. In this Research Note, we show that the small orbital eccentricities of HAT-P-7b and Kepler-13Ab cannot generate the odd harmonics observed in these systems.

\begin{figure}[htb]
\begin{center}
\includegraphics[scale=0.5,angle=0]{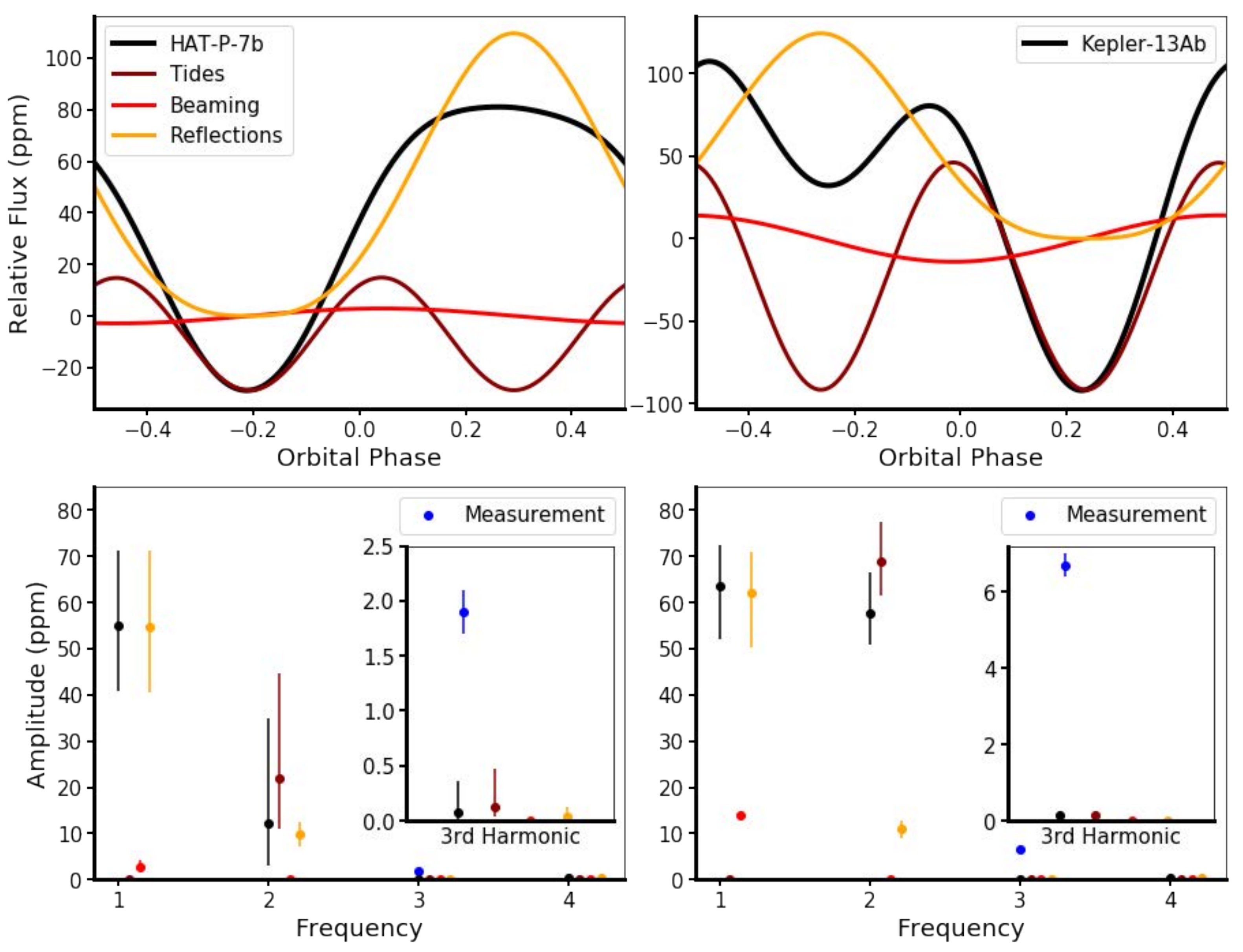}
\caption{The top panels show orbital lightcurves generated using the \emph{Out of Transit} code with the transits and eclipses removed for clarity \citep{2018arXiv180307078P}. The bottom panels show the Fourier amplitudes of the lightcurves. The sum of the effects from tides, beaming and reflection is shown in black.  The \emph{Kepler} measurements of \cite{2015ApJ...804..150E} are shown in blue.  The measured amplitudes are 5$\sigma$ and 22$\sigma$ greater than predicted for HAT-P-7 and Kepler-13Ab, respectively. Tidal deformation due to an eccentric planet cannot be responsible for these odd modes.
\label{fig:1}}
\end{center}
\end{figure}

\vspace{0.3cm}
We use the \emph{Out of Transit} code \citep[OoT;][]{2018arXiv180305917P,2018arXiv180307078P} to simulate light curves for HAT-P-7 and Kepler-13A including reflection, Doppler beaming, and time-dependent tidal distortion using system parameters from \cite{2016ApJ...823..122W} and \cite{2015ApJ...804..150E}, respectively. 
We numerically integrate the model light curves to obtain the corresponding Fourier amplitudes (both shown in Figure~\ref{fig:1}).  
The 1$\sigma$ uncertainty intervals on the predicted amplitudes are evaluated using a Monte Carlo over the mass and radius of the host star and planet, semi-major axis, eccentricity, inclination, argument of periapsis, and geometric albedo based on the published values and their uncertainties (for parameters with asymmetric uncertainties a two-piece Gaussian distribution was used). 
For Kepler-13Ab, we adopt a geometric albedo of 0.33$^{+0.04}_{-0.06}$ \citep{2014ApJ...788...92S} while for HAT-P-7b we use 0.3.\footnote{Our chosen albedo value for HAT-P-7b is an order of magnitude greater than the albedo reported by \cite{2013ApJ...772...51E}, but is required to match the \emph{Kepler} observations of this system: since OoT neglects thermal emission from the planet, geometric albedo is the only lever to increase the amplitude of variations at the orbital frequency. This subtlety is immaterial to the current analysis, however: neither reflectance nor thermal emission from a static planetary map can contribute odd harmonics at the levels reported for these planets, as shown by \cite{2017MNRAS.467..747C} and illustrated in the insets of Figure~\ref{fig:1}. Moreover, the time variations in albedo reported for HAT-P-7b \citep{2016NatAs...1E...4A} are far too slow to account for the observed odd mode amplitude \citep{2017MNRAS.467..747C}.}   

The models in Figure~\ref{fig:1} are predictions based on published system parameters, rather than fits to the observed phase curves; nonetheless, they match the fundamental mode ($\sim$reflection+beaming) and second mode ($\sim$ellipsoidal variations) reasonably well. Our Monte Carlo analysis produces an amplitude for the third mode of 
$0.14 ^{+0.03}_{-0.04}$~ppm ($1\sigma$ uncertainty) for Kepler-13Ab
and a $3\sigma$ upper limit of 
$0.92$~ppm for HAT-P-7b. Those values are much lower than the values reported by \cite{2015ApJ...804..150E}: $1.9 \pm 0.2$~ppm for HAT-P-7b and $6.7 \pm 0.3$~ppm for Kepler-13Ab.  
While increasing the $\beta$ tidal parameter in OoT can successfully predict the observed 3$^{\rm rd}$ mode amplitude, the second mode's amplitude is increased by the same factor and becomes discrepant with the observations. Orbital eccentricity increases the odd harmonics without greatly increasing the ellipsoidal contribution, but both planets are on nearly circular orbits: 
$e_{\rm H7}=0.0016^{+0.0034}_{-0.0010}$ \citep{2016ApJ...823..122W} and 
$e_{\rm K13}=0.00064^{+0.00012}_{-0.00016}$ \citep{2015ApJ...804..150E}.
We therefore conclude that the odd harmonics observed in these systems are not due to orbital eccentricity.  \cite{2015ApJ...804..150E} hypothesized that the odd modes could be due to tidal distortion of the star if its spin is misaligned with the system's orbital motion---as is the case in both of these systems \citep[][]{2014PASJ...66...94B, 2018AJ....155...13H}---but this mechanism has yet to be verified theoretically or numerically.   

\vspace{0.2cm}
The authors thank Z.~Penoyre for sharing OoT and thank he, N.~Stone, and L.~Esteves for feedback on this manuscript, as well as E.~Pass for her two-piece normal distribution Python function. 

\end{document}